\documentclass[prd,showpacs,preprintnumbers,nofootinbib,aps]{revtex4}
\usepackage{subeqnarray}
\usepackage{epsfig,amsmath,amssymb}
\usepackage{mathrsfs}
\usepackage[usenames,dvipsnames]{color}
\usepackage[pagebackref=true, colorlinks=true]{hyperref}
\definecolor{redish}{rgb}{0.7,0.2,0.0}  
\definecolor{bluish}{rgb}{0.2,0.5,0.8}
\hypersetup{linkcolor=redish,          
                  citecolor=blue,        
                  filecolor=magenta,      
                  urlcolor=bluish}          
\DeclareFontFamily{U}{rsfs}{}         
\DeclareFontShape{U}{rsfs}{m}{n}{<5> rsfs5 <6><7> rsfs7          %
  <8><9><10><10.95><12><14.4><17.28><20.74><24.88> rsfs10}{}     %
\DeclareMathAlphabet{\mathfs}{U}{rsfs}{m}{n}                     %
\newcommand{\mfs}[1]{\mathfs {#1}}                               %

\newcommand{\ba}{\nopagebreak[3]\begin{eqnarray}}
\newcommand{\ea}{\end{eqnarray}}
\newcommand{\bii}{\begin{itemize}}
\newcommand{\eii}{\end{itemize}}
\newcommand{\nn}{\nonumber}

\newcommand{\sO}{{\mfs O}}

\newcommand{\f}{\frac}

\def \d{\delta}
\def \b{\beta}

\def \g{\gamma}
\def \lp{\ell_p}

\def \lm{\lambda}

\def \s{\sigma}

\def \H{\mathcal{H}}
\def \P{\mathcal{P}}

\def \mic{{\it microcanonical}}
\def \sj{ s^{\star}_j}
\def \3{\sqrt{3}}
\begin{document}

\title{The Microcanonical Entropy of Quantum Isolated Horizon, `quantum hair' $N$ and the Barbero-Immirzi parameter fixation }
\author{Abhishek Majhi}%
 \email{abhishek.majhi@saha.ac.in}
\affiliation{Saha Institute of Nuclear Physics,\\Kolkata 700064, India}%
\pacs{04.70.Dy, 04.60.Pp, 03.65.Aa, 03.67.-a, 05.30.-d, 05.30.Ch}

\begin{abstract}
{\it If} the total number of punctures($N$) of a quantum isolated horizon is considered to be a macroscopic parameter alongside the Chern-Simons level($k$) or equivalently classical area$(A_{cl})$ a strict analysis of the {\mic} ensemble reveals that the {\it microcanonical} entropy has the form $S_{MC}=A_{cl}/4\lp^2+N\s(\g)$, only for values of the Barbero-Immirzi(BI) parameter$(\g)$ greater than a certain number. It is argued that the term $N\s(\g)$ must be negative definite, which leads to the bound on the BI parameter. 
\end{abstract}

\maketitle
\section{Introduction}\label{sec1}
In {\it loop quantum gravity}({\bf LQG}) framework, the  bulk spin network edges intersecting the classical {\it Isolated Horizon}({\bf IH})\cite{ih1,ih2,ih3,ih4,ih5} at the so called {\it punctures}, depict the {\it Quantum Isolated Horizon}({\bf QIH}) \cite{qg1,qg2}. The Hilbert space of the QIH \cite{qg1,qg2} is spanned by the states of puncture(source) coupled {\it Chern-Simons}({\bf CS}) theory, the level of the CS theory being given by $k\equiv A_{cl}/4\pi\g\lp^2$ where $A_{cl}$ is the area of the classical IH, $\g$ is the {\it Barbero-Immirzi}({\bf BI}) parameter and $\lp$ is the Planck length.  Each of the punctures is associated with an SU(2) spin deposited by the piercing edges of the bulk spin network which can have a maximum value of $k/2$ \cite{km1}. The kinematical Hilbert space of the QIH  provides a self contained platform for the direct application of statistical mechanics in view of exploring the thermodynamics where the only external input needed to fix the $\g$-ambiguity of the underlying theory is the {\it Bekenstein-Hawking area law}({\bf BHAL})\cite{bhal}.

\par
In  \cite{GP}, besides the classical area$(A_{cl})$ or equivalently $(k)$ the CS level, the total number of punctures $(N)$ on the QIH is considered as a macroscopic parameter to define the microcanonical ensemble in contrast to all previous entropy calculations where {\it only} $A_{cl}$ or $k$ is used to define the {\mic} ensemble \cite{qg1,qg2,km2,sig,enp,bunch1,bunch2,dole,mei,gm1,gm2,gm3,gm4}.
  The concept of `quantum hair' $N$ introduced in \cite{GP} has been adored by several authors resulting in a series of papers where it has been naively overlooked that as a consequence of the introduction of `quantum hair' $N$, $\g$ can neither be uniquely determined as in \cite{qg1,qg2,km2,sig,bunch1,bunch2,dole,mei,gm1,gm2,gm3,gm4}, nor it is a free parameter as opposed to the claim of \cite{GP}. In fact, a careful analysis of the derivation of the {\mic} entropy of QIH for given $k$ and $N$ reveals that $\g>0.191$ for the {\mic} entropy to be given by
\ba
S_{MC}=\f{A_{cl}}{4\lp^2} + N\s(\g)\label{SMC.}
\ea
reported in \cite{GP}. The bound on $\g$ follows from the fact that the term $N\s(\g)$ must be a negative definite quantity. 
It is worth emphasizing that in this work we neither support nor contradict the idea of `quantum hair' $N$ proposed in \cite{GP}, but only discuss its implications in the context of {\mic} entropy calculation {\it if} one accepts the idea and considers $N$ to be an additional macroscopic parameter for a QIH, alongside $k$ or $A_{cl}$. 
\par
The contents of the paper can be debriefed as follows.
In section(\ref{sec2}), an outline of the calculation of the {\mic} entropy of a QIH for given $k$ and $N$ is provided, followed by a detailed discussion on the role of the Lagrange multipliers in the {\mic} ensemble. In section(\ref{sec3}), the arguments in favor of the  negativity of the term $N\s(\g)$ are explained in details. The fact that $\s(\g)$ is negative leads to a bound on the allowed values of $\g$. The paper ends with section(\ref{sec4}) which contains the conclusions.

\section{Entropy of QIH and the Lagrange multipliers in {\it microcanonical} ensemble}\label{sec2}
Following \cite{GP}, {\it considering} $N$ as a macroscopic parameter for a QIH alongside $k$, we shall first compute the {\mic} entropy of a QIH for given $k$ and $N$ \cite{me2}, using the standard statistical method of most probable distribution\cite{lapa}. Although this calculation has been exhaustively carried out in \cite{me2} but it will be still convenient for the reader to have the outline of the derivation reiterated here so as to have a consistent understanding of the picture.

\par
An area eigenstate of a QIH is given by  an SU(2) spin configuration given by the set $\{s_j\}$ \cite{me2}, where $s_j$ is the number of punctures with spin value $j$. The number of microstates for a particular spin configuration $\left\{s_j\right\}$ in the SU(2) approach\cite{km1,bunch1,bunch2}  is given by
$\Omega[\left\{s_j\right\}]=\frac{N!}{\prod_{j}s_j!}\frac{2}{k+2}\sum^{k+1}_{a=1}\sin^2\frac{a\pi}{k+2}\prod_{j}\left\{\frac{\sin\frac{a\pi(2j+1)}{k+2}}{\sin\frac{a\pi}{k+2}}\right\}^{s_j}$. The summation over $a$ can be replaced by an integration in the large $k$ limit. The integration can be easily evaluated by saddle point approximation. The zeroth order term yields $g[\{s_j\}]\sim (2j+1)^{s_j}$ which is used in \cite{GP}. We shall consider only this result which is sufficient as far as the subject matter of this paper is concerned and hence write
$\Omega[\left\{s_j\right\}]=C\f{N!}{\prod_{j}s_j!}(2j+1)^{s_j}$
where $C$ is a constant in the zeroth order approximation. Considering the quadratic fluctuations in $a$ modifies the above result in such a way that it yields a logarithmic correction in the entropy with a universal coefficient of $-\f{3}{2}$. The details of these calculations can be found in \cite{me2}.

\par
\begin{figure}[hbt]
\begin{center}
\includegraphics[scale=0.85]{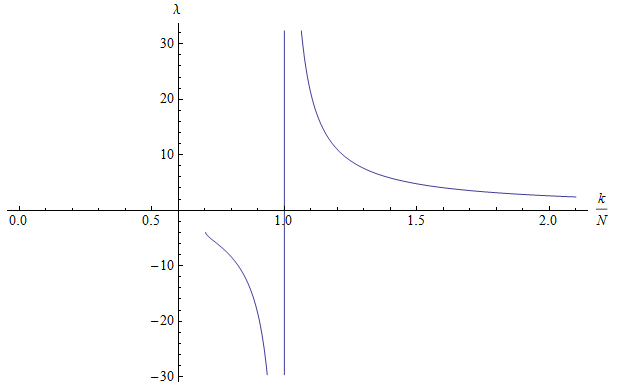}
\caption{\label{fig:lmkn}The above plot shows the variation of $\lm$ with $k/N$ for the solution of $\lm$ with the `+' sign in eq.(\ref{lmsol})}. It is quite clear that the value of $\lm$ has a discontinuity at $k/N=1$ and has positive values only for $k/N>1$.
\end{center}
\end{figure}

{\it If} we define a {\it microcanonical} ensemble of QIHs for given values of $N$ and $k$, the spin configurations $\{s_j\}$ must obey the following constraints
\begin{subeqnarray}\label{con}
&&{\cal C}_1 : \sum^{k/2}_{j=1/2} s_j = N \slabel{con1} \\
&&{\cal C}_2 : \sum^{k/2}_{j=1/2} s_j \sqrt{j(j+1)} = k/2\slabel{con2} 
\end{subeqnarray}
Variation of $\log \Omega[\left\{s_j\right\}]$ with respect to $s_j$, subject to the constraints ${\cal C}_1$ and ${\cal C}_2$, yields the most probable distribution $\sj$, which maximizes the entropy of the QIH and can be found to be
\ba
\sj=N(2j+1)e^{-\lm\sqrt{j(j+1)}-\s}\label{dc}					
\ea
where $\s$ and $\lm$ are the Lagrange multipliers for ${\cal C}_1$ and ${\cal C}_2$ respectively. Using the dominant distribution $s_j^{\star}$ to satisfy the two constraints (\ref{con1}) and (\ref{con2}), considering the large $k$ limit where one can safely replace\footnote{The approximation done by replacement of the summation by an integration is analogous to the case of the famous Stirling's approximation.  The easiest method to arrive at the result is the following : 
$\lim_{N\to\infty}\log N!= \lim_{N\to\infty}\sum_1^N\log n\simeq \int_1^N\log x ~dx\simeq N\log N - N$. This is available in standard textbooks of statistical mechanics e.g.  the second textbook referred to in \cite{lapa}.}
 $\lim_{k\to\infty}\sum_{j=1/2}^{k/2}$ by $\int_{1/2}^{\infty} dj$, 
it is straightforward to carry out the integrations to obtain
\ba
e^{\s}&=&\f{2}{\lm^2}\left(1+\f{\3}{2}\lm\right)e^{-\f{\3}{~2}\lm}\label{rel1}\\
\f{k}{N}&=&1+\f{2}{\lm}+\f{4}{\lm(\3\lm+2)}\label{rel2}
\ea
Since we are dealing with the {\mic} ensemble, $k$ and $N$ are the given quantities and   the Lagrange multipliers $\lm$ and $\s$ can be obtained as the solutions of the equations (\ref{rel1}) and (\ref{rel2}).
 It can be checked explicitly in the following way. Eq.(\ref{rel2}) is actually a cubic equation in $\lm$ written as
\ba
\lm\left[\sqrt{3}\left(k/N-1\right)\lm^2+\left(k/N-1-\sqrt 3\right)\lm-8\right]=0
\ea
Excluding the trivial root $\lm=0$ of the above equation for obvious reasons\footnote{$\lm=0$ leads to $\s\to\infty$ which will yield infinite entropy.}, the other two nontrivial roots of the above equation are given by
\ba
\lm=\f{1}{\sqrt{3}(k/N-1)}\left[\left(\sqrt 3+1-k/N\right)\pm\sqrt{k^2/N^2+\left(6\sqrt 3-2\right)k/N+\left(4-6\sqrt 3\right)}\right]\label{lmsol}
\ea
of which we shall again exclude the one with the `$-$' sign because it will yield negative values of $\lm$ for all $k/N>0$ and hence leading to negative values of BI parameter (it will be clear shortly). Hence, we shall consider only the one with the `$+$' sign as this will only give the positive values of $\lm$ for $k/N> 1$. To see this, one can plot\footnote{All the graph plots shown in this paper are performed with MATHEMATICA.} $\lm$ as a function of $k/N$ considering the expression with the `+' sign. The resulting graph is shown in in FIG.(\ref{fig:lmkn}). Using the desired solution of $\lm$ as a function of $k/N$ in eq.(\ref{rel1}) it is trivial to obtain $\s$ as a function of $k/N$. Hence, in the {\mic} ensemble, $\lm\equiv\lm(k/N)$ and $\s\equiv\s(k/N)$ are functions of $k$ and $N$. Eq.(\ref{rel2}) can be considered to be the equation of state relating $\lm,k$ and $N$ only at the equilibrium and hence can be attained only after finding the most probable distribution giving the equilibrium configuration. It should be noted that {\it there is no freedom to choose $\lm$ in the {\mic} ensemble.}

Now, the {\mic} entropy is given by $S_{MC}=\log\sum_{\{s_j\}}\Omega[\left\{s_j\right\}]$. Taking into account that the dominant contribution comes from the most probable configuration $\{\sj\}$ which maximizes the entropy and taking the limit $N,\sj\to\infty$ so as to apply the Stirling approximation, one can  calculate the {\it microcanonical} entropy as
\ba
S_{MC}&\simeq&\lim_{N,\sj\rightarrow\infty}\log \Omega\left[\left\{\sj\right\}\right]\nn\\
&=&\lm(k/N) k/2+ \s(k/N) N\label{smc}
\ea
where one has to use also the eq.(\ref{con1}) and eq.(\ref{con2}). 
Thus, once we define the {\mic} ensemble of QIHs by giving $k$ and $N$, the {\mic} entropy is completely known and given by eq.(\ref{smc})\footnote{The above scenario is analogous to the case of an ideal gas whose equation of state is  given by $E=\f{3}{2}NT$(considering Boltzmann constant to be unity and the meaning of $E,N$ and $T$ are obvious). Since the {\mic} ensemble is defined by given values of $E$ and $N$, $T$ is a derived quantity and should be viewed as $T\equiv T(E,N)$. Thus the {\mic} entropy of an ideal gas for given $E$ and $N$ must be written as 
\ba
S_{MC}=\b(E,N)E+N\alpha(E,N)\label{comp1}
\ea
where $\b=1/T$ and $\alpha$ are the Lagrange multipliers solved for given $E$ and $N$ \cite{lapa}. It is only in the {\it canonical} ensemble one can say that $T$ can be chosen because the ensemble is defined by specifying the equilibrium temperature $(T)$ and the total number of particles$(N)$. In this case $T$ and $N$ are the given quantities and $E\equiv E(T,N)$ becomes a derived quantity i.e. we {\it calculate} the mean energy of the system at a desired temperature and for a desired number of particles\cite{lapa}. The {\it canonical} entropy of an ideal gas for given $T$(or equivalently $\b$) and $N$ must be written as
\ba
S_{C}=\b E(\b, N)+N\alpha(\b, N)\label{comp2}
\ea
It is a very crucial point to be noted that even though the structure of the thermodynamic equations, such as the form of entropy in eq.(\ref{comp1}) and eq.(\ref{comp2}), the equation of state, etc. are {\it independent} of the ensemble we use, the point which is often overlooked is that the roles of the parameters $(E,N,T)$ indeed change with the ensemble as discussed above.}.

\par
Now, we can write eq.(\ref{smc}), by replacing $k$ with $A_{cl}/4\pi\g\lp^2$, in the following form
 \ba
 S_{MC}=\f{\lm(k/N)}{2\pi\g}\f{A_{cl}}{4\lp^2} + \s(k/N) N  \label{SMC}
 \ea 
Our goal is to obtain the form of the {\mic} entropy given by the expression (\ref{SMC.}) from the expression (\ref{SMC}). Usually, in calculation of entropy only for fixed $k$ or $A_{cl}$(e.g. see \cite{sig}) (i.e. the scenario which appears by putting $\s=0$ in the present case), a fixed numerical value of $\g$ is determined by demanding the BHAL. In that case $\lm$ comes out to be a number and $\g$ is chosen to get the desired result, which is consistent with the fact that $\g$ has to have a specific numerical value so as to have an unambiguous LQG theory. But, in the present scenario with an additional macroscopic parameter $N$, $\lm$ is a function of $k/N$. Hence, there is no other way than to accept that $\g=\lm(k/N)/2\pi$, a function of $k$ and $N$ \footnote{Let us define the {\mic} ensemble by assigning values $k=k_1$ and $N=N_1$, for which we have $\lm=\lm(k_1/N_1)\equiv\lm_1$. Now, we claim that $\g=\lm_1/2\pi=\g_1$(say) so that we can obtain the first term of (\ref{SMC}) to be given by $A_{cl1}/4\lp^2$, where $A_{cl1}=4\pi\g_1 k_1\lp^2$. Similarly, one can make another choice $k=k_2$ and $N=N_2$, such that $(k_1/N_1)\neq(k_2/N_2)$, for which there exists a corresponding $\g_2$ and $A_{cl2}$ so as to obtain the first term of (\ref{SMC.}) to be $A_{cl2}/4\lp^2$. To be precise for every such choice of $k/N$ there exists an unique value of $\g$, given by $\lm(k/N)/2\pi$, which results in the {\mic} entropy given by (\ref{SMC.}).} 
, so as to obtain the {\it microcanonical} entropy  of the form given by the expression (\ref{SMC.}).

Hence, for the {\mic} entropy to be given by (\ref{SMC.}), we must have $\g=\lm(k/N)/2\pi$ and there is no way one can obtain a specific universal value of $\g$ and it is indeed a {\it function} of $k/N$ in this scenario where $N$ is considered to be a macroscopic parameter for a QIH alongside $k$ \footnote{Here one may wonder if this problematic fixation of $\g$ is a result of considering $k$ to be a macroscopic parameter preferred to $A_{cl}$. But one can remain assured that this is not the actual reason and to get convinced (s)he may check by repeating this whole calculation by fixing $A_{cl}$ instead of $k$, alongside $N$ to define the {\mic} ensemble.}. But that does not mean that $\g$ is a free parameter, unlike what has been claimed in \cite{GP}. There is a bound on the allowed values of $\g$ which follows from strong arguments to be explained in details shortly. By now we can conclude that for each value of $k/N$, there exists a unique value of $\g$ for which the {\mic} entropy takes the form of expression (\ref{SMC.})
reported in \cite{GP}. The  allowed values of $\g$ is restricted by the bound : $\g>0.191 $. This bound obviously needs an explanation which is the subject matter of the next section.

\section{ Bound on $\g$}\label{sec3}
In this section we shall argue from two different viewpoints that the term $N\s(\g)$ should be a negative definite quantity from which the bound on $\g$ will follow. First of all we shall explain this by looking at the kinematical Hilbert space structure of the QIH, which is usually studied for calculating black hole entropy in LQG framework. The second argument originates from the comparison of the entropy measured by a local stationary (with respect to the horizon) observer with the one measured by the observer at asymptotic infinity for the same QIH. The section ends with a quantitative estimate of the bound on $\g$.

\subsection{Constrained kinematical Hilbert space}
Imposing constraints on a system implies availability of more information about that system. Since, entropy is a measure of unavailability of information about a system\cite{bril}, thus imposition of more constraints will result in decrement of the entropy. This is what happens also in the case of black hole entropy which, in the LQG framework, is calculated by taking the logarithm of the dimensionality of the associated kinematical Hilbert space. As far as the full kinematical Hilbert space of a QIH is concerned\cite{qg1,qg2,enp}, it is interesting to note that there is actually a {\it sum over all possible sets of punctures} which encodes the information that the full Hilbert space of the QIH takes into account all possible values of $N$ compatible with a given $k$ :
\ba
\H_{QIH}^{k}&=&\bigoplus_{\{\P\}}\text{Inv} \left(\bigotimes_{l=1}^N \H_{j_l}\right)\label{hsqih}
\ea
where ${\{\P\}}\equiv {N ; \f{1}{2}\leq j_l\leq\f{k}{2}\forall l\in[1,N] \ni \sum_{l=1}^N\sqrt{j_l(j_l+1)}=\f{k}{2}\pm\sO(\f{1}{8\pi\g})} $ and `Inv' stands for the gauge invariance. Prior to the advent of the concept of quantum hair in \cite{GP}, the dimensionality of this full Hilbert space was considered which gave the total number of horizon microstates for a given $k$ and the entropy used to come out to be the BHAL for a unique value of $\g$\cite{qg1,qg2,dole,mei}.
Now, {\it if} one considers $N$ as an independent macroscopic parameter other than $A_{cl}$ or $k$ and $N$ is specified to define the {\mic} ensemble, then  the resulting {\mic} entropy will be that of a fixed-$N$ subspace of the full kinematical Hilbert space. Since the dimensionality of this subspace is bound to be less than that of the full kinematical Hilbert space, the resulting entropy must be less than the BHAL i.e. the term $N\s(\g)$ should only appear as a negative term so as to lower the entropy below BHAL. 
\\
{\it Remarks :} This is the same reason why the logarithmic correction to the BHAL from quantum geometry  comes with a negative sign which has been explained in details in the {\it appendix} of this paper.

\subsection{Local vs Asymptotic Views}

As has been clearly explained in \cite{GP} that,  the proposal of the quantum hair $N$ has been given from the local stationary observer perspective i.e. an observer at a proper distance of few Planck lengths from the horizon and stationary with respect to the horizon, will realize the existence of the quantum hair $N$. It implies that {\it only} the local observer can treat the total number of punctures $N$ as a macroscopic thermodynamic parameter, but the asymptotic observer does not realize the existence of this quantum hair $N$. The fluctuations of $N$ appear to the asymptotic observer as small quantum fluctuations, which has no effect on the thermodynamics at asymptotic infinity, as opposed to the local observer who can treat $N$ as a macroscopic thermodynamic parameter because the fluctuations of $N$ indeed appear to the local observer as particle like excitations on the horizon. This is why the chemical potential conjugate to $N$ which exists for the local observer, must vanish at asymptotic infinity \cite{GP}.  For the {\it same} system i.e. the QIH, there are two observers and hence two different observations. The local observer gives us a fine grained view whereas the asymptotic observer gives us a coarse grained view of the {\it same} system. The local observer has an access to larger amount of information than the asymptotic observer has about the {\it same} system, the QIH and that is why $N$ can be treated as a macroscopic parameter only by the local observer and not by the asymptotic observer.  Thus the entropy of the QIH measured by the local observer must be less than the entropy of the same QIH measured by the asymptotic observer. Hence, the $N\s(\g)$ term, which is seen by the local observer only, must be negative definite i.e. $\s(\g)<0$.
\\
{\it Remarks :} One should note that this above argument stands only because we know or accept that the asymptotic observer must observe the BHAL. Due to the presence of this `reference' measurement we could argue that the entropy of the QIH measured by the local observer must be less than this `reference' BHAL. In general gas thermodynamics  no such difference in observations is made and the $N\s$ like term that appears there can be anything : positive, negative or zero. This is a crucial point to be noted.


\subsection{An estimate of the bound on $\g$}
Following the above qualitative arguments in favour of the boundedness of $\g$ resulting from the bound $\s(\g)<0$, it is the turn to show off a quantitative analysis on behalf of the claim. From eq.(\ref{rel1}) it is quite easy to get an estimate of the bound on $\g$. If one plots $e^{\s}$ as a function of $\lm$, it is seen that the value of $e^{\s}$ falls below $1$ i.e. $\s$ becomes negative for $\lm>1.200$. Now, following the previous arguments regarding the fixation of $\g$, one can obtain the bound on $\g$ by dividing the allowed range of $\lm$ by $2\pi$, which results in $\g>0.191$.
\begin{figure}[hbt]
\begin{center}
\includegraphics[scale=1.00]{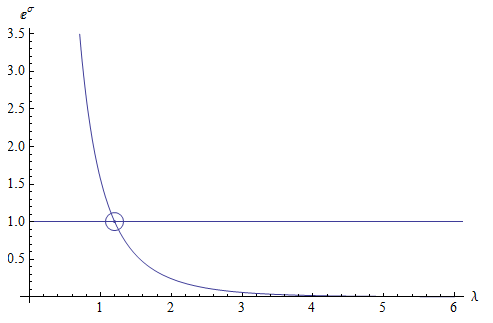}
\caption{\label{fig:gma} In the above plot of $e^{\s}$ as a function of $\lm$, the coordinates of the marked point in the graph are $(1.200,1.000)$. Therefore, one can conclude that $e^{\s}<1\Rightarrow \s<0$ for $\lm>1.200$. Since $\g=\lm(k/N)/2\pi$, we obtain the required bound on the BI parameter i.e. $\g>0.191$. }
\end{center}
\end{figure}

\section{Conclusion}\label{sec4}
To conclude, it is worth emphasizing that nowhere in this work we either justify or contradict the idea of considering $N$ as a macroscopic parameter for a QIH put forward in \cite{GP}. This work is presented only to assert that {\it if} one accepts $N$ as a `quantum hair' of  QIH and define the microcanonical ensemble for given $k$ or $A_{cl}$ and $N$, then to obtain the {\mic} entropy of a QIH given by eq.(\ref{SMC.}), which was first reported in \cite{GP}, we must have $\g>0.191$. This is not quite in agreement with \cite{GP}, according to which $\g$ is a {\it free} parameter i.e. it can take any value.  Moreover, there is no way one can obtain a unique value of $\g$ in this particular scenario. The BI parameter, $\g$, being of utmost importance in LQG, any further work following the idea of `quantum hair' $N$ must be performed with careful attention to the bound on $\g$ which has been somehow overlooked in \cite{GP}. It is to be noted that the bound  on $\g$ may be calculated more precisely by numerical methods, but the motto of this paper is to catch the essence of the boundedness of $\g$,  which is the most essential physics content in the present context and it is not the mathematical accuracy of the numbers that we are after. It should be reminded that the pivotal point of this work consists of the arguments in favor of the condition $N\s(\g)<0$ which results from viewing the problem from a very different perspective. In general, while studying the thermodynamics of a system, we do not talk precisely about the observer and all the measurements made are considered to be unique. But the topic of black hole thermodynamics which is related to general relativity, the observer must play a crucial role in the measurements. Since we have accepted by heart and soul that the observer at asymptotic infinity will measure the entropy to be nothing other than the BHAL, then, whatever observer and corresponding measurement we consider, there has to be a consistency with the known measured value at asymptotic infinity. Our arguments simply stand on this ground. If there {\it were} no BHAL, then we could not have presented any of our arguments in favor of $N\s(\g)<0$. Then, it could have had arbitrary sign and $\g$ would have been a free parameter. 

Finally, we may mention that considerations related to the Entropy Bound \cite{bek}, as `covariant'-ized in \cite{bou} and sharpened within LQG in \cite{srp}, places our qualitative arguments in favor of $\s(\g)<0$ on a stronger footing. As it has been pointed out in course of the presentation of our arguments, the idea of quantum hair $(N)$ is an observer dependent notion\cite{GP} i.e. $N$ can be considered as a macroscopic thermodynamic variable only by a local observer very close to the horizon and there is no such notion of quantum hair for an asymptotic observer. Now, it is already known in the literature that there is a covariant (observer independent) entropy bound associated with a closed two-surface\cite{bek,bou} which has been proved quantum geometrically in \cite{srp} leading to a tighter bound for a QIH. For a closed spatial two-surface of area $A_{cl}$ the maximum associated entropy can only be $A_{cl}/4\lp^2$ (ignoring the logarithmic correction\cite{srp}). Since the horizon entropy is nothing but the entropy associated with the closed two-surface cross-sections, it is evident that whatever observer dependent entropy one can calculate,  cannot be greater than $A_{cl}/4\lp^2$. Hence, the observer dependent notion of quantum hair $N$ can only give rise to a negative contribution to the entropy and thus there is no other choice than to impose the condition $\s(\g)<0$.

{\it Note Added }: It is worth mentioning that it is indeed possible to render the BI parameter to be {\it free} even in this setup where $N$ is considered as an independent macroscopic parameter alongside $k$, if one naively demands that the entropy be given by the BHAL only. It trivially follows from eq.(\ref{smc}), which can as well be written as $S_{MC}=\tilde\lm(k/N)k/2$, where $\tilde \lm=\lm-\left(\s/\f{d\s}{d\lm}\right)$. Now, requiring that the BHAL must follow one has to fit $\g=\tilde\lm(k/N)/2\pi$. As there is no additional term to the BHAL, therefore there is no question of any further arguments. Study of the function $\tilde\lm$ reveals that it can take values from 0 to $\infty$ thus resulting in no bound on $\g$. One can look into \cite{me2} for details on this issue. The author thanks one of the referees for suggesting to comment on this issue which is quite important in the present context.

The idea of a complex BI parameter arising from a  formulation of general relativity based on the self-dual Sen-Ashtekar connection is an intriguing possibility. However, such a   formulation necessarily deals with a complex configuration space which  leads to mathematical difficulties when quantization is  attempted \cite{aslew}. 
As far as the black hole entropy computation is concerned, it may be noted that the comparison of the QIH entropy, derived from a purely quantum statistical calculation,  with the semiclassical BHAL may be fraught with a slight danger since there is as yet no complete semiclassical formulation derived from the coherent states of LQG. There is indeed the need for an appropriate effective action of the theory which may result in a renormalized BI parameter. The situation is reminiscent of the $\theta$-parameter in QCD, because of the topological character of the BI parameter \cite{sandipan}. In QCD, too, the comparison of phenomenological results based on $\theta$-vacua with observations {\it assumes} an effective `renormalized' $\theta$ parameter. The author  thanks the  Referee for pointing out ref.\cite{mg,nb}.


\section{appendix}
~\\
{\bf Quantum Information and Constraints :} If one considers the SU(2) and U(1) microstates of the CS theory (coupled to punctures) and find out the entropy for fixed area($A_{cl}$) of the QIH and considering number of punctures $N$ to be arbitrary (which was usually done before \cite{GP}), it is well known that the logarithmic corrections are $-3/2\log A_{cl}$ and $-1/2\log A_{cl}$ respectively, alongside the BHAL. As has been correctly pointed out in \cite{km2} that the availability of quantum information results in the {\it negative} correction, thus reducing the entropy from the  BHAL resulting from the semiclassical information theory\cite{bhal}. Going one step further, one can also realize this effect by looking at the difference in quantum log corrections from SU(2) and U(1) theory. The SU(2) entropy is less than the U(1) entropy. This is simply due to the fact that the number of microstates in the U(1) CS theory is reduced by the imposition two more constraints (more information) to give the number of microstates in the SU(2) theory. It follows from the following expression for the SU(2) microstates \cite{km1,sig}
\begin{widetext}
\ba
\Omega_{SU(2)}[j_1,j_2, \cdots, j_N]&=&\sum_{m_1=-j_1}^{j_1}\cdots\sum_{m_N=-j_N}^{j_N}\left[~\underbrace{\d_{\left(\sum_{p=1}^{N}m_p\right),0}}_{\text{U(1) microstates}}~~~\underbrace{-~~\f{1}{2}~\d_{\left(\sum_{p=1}^{N}m_p\right),1}~-~~\f{1}{2}~\d_{\left(\sum_{p=1}^{N}m_p\right),-1}}_{\text{Constraints that cancel the unphysical over counting}}~\right]\nn
\ea
\end{widetext}
Straightforward explicit calculations have been done in \cite{sig,me2} which clearly  show that if one just considers the U(1) term in the above expression, the calculation yields $-1/2\log A_{cl}$ and further consideration of the last two terms, which nullifies the unphysical over counting of microstates of U(1) theory \cite{enp}, results in an additional $-\log A_{cl}$. The complete result will be the $-3/2\log A_{cl}$ added to the BHAL. Thus, the entropy is decreased as the effect of availability of quantum information of the system\cite{bril} provided by the constraints.

\section{Acknowledgments}
 As always, I am grateful to my supervisor Prof. Parthasarathi Majumdar, for his priceless advices and encouragement, alongside many illuminating discussions regarding my work and also for helping me to improve the draft.  Discussions with Prof. Daniele Oriti and Daniele Pranzetti are gratefully acknowledged. Part of the work was done during a visit to the Albert Einstein Institute, Potsdam, Germany. My sincere thanks go to the referees whose suggestions and criticisms have resulted in potential improvement of the manuscript. I sincerely acknowledge the financial support provided by the Department of Science and Technology of India to carry out my research work.

\end{document}